# Integral Equation Analysis of Plane Wave Scattering by Coplanar Graphene-Strip Gratings in the THz Range

Olga V. Shapoval, Juan Sebastian Gomez-Diaz, Julien Perruisseau-Carrier, Juan R. Mosig, and Alexander I. Nosich

*Abstract*— **The plane wave scattering and absorption by finite and infinite gratings of free-space standing infinitely long graphene strips are studied in the THz range. A novel numerical approach, based on graphene surface impedance, hyper-singular integral equations, and the Nystrom method, is proposed. This technique guarantees fast convergence and controlled accuracy of computations. Reflectance, transmittance, and absorbance are carefully studied as a function of graphene and grating parameters, revealing the presence of surface plasmon resonances. Specifically, larger graphene relaxation times increases the number of resonances in the THz range, leading to higher wave transmittance due to the reduced losses; on the other hand an increase of graphene chemical potential up-shifts the frequency of plasmon resonances. It is also shown that a relatively low number of graphene strips (>10) are able to reproduce Rayleigh anomalies. These features make graphene strips good candidates for many applications, including tunable absorbers and frequency selective surfaces.**

## I. INTRODUCTION

GRAPHENE monolayers are electrically infinitesimal thin (single-atom) and display a rather good electron conductivity $\sigma$ that mainly depends on frequency, temperature, electron relaxation time and chemical doping [1-10]. In addition, one of the most promising features of graphene as compared with metals is the opportunity to modify its conductivity by applying and external electrostatic biasing field, which modify graphene chemical potential [1-2]. This can be easily implemented for instance by including an extremely thin polysilicon layer below the dielectric which supports graphene, and applying a DC bias between these two layers (see [1,2,8]). Graphene is very interesting in view of possibility of strong interaction with electromagnetic waves in the THz frequency range [3,4]. Indeed, it is able to support delocalized surface-plasmon waves at frequencies two orders of magnitude lower than the noble metals [3].

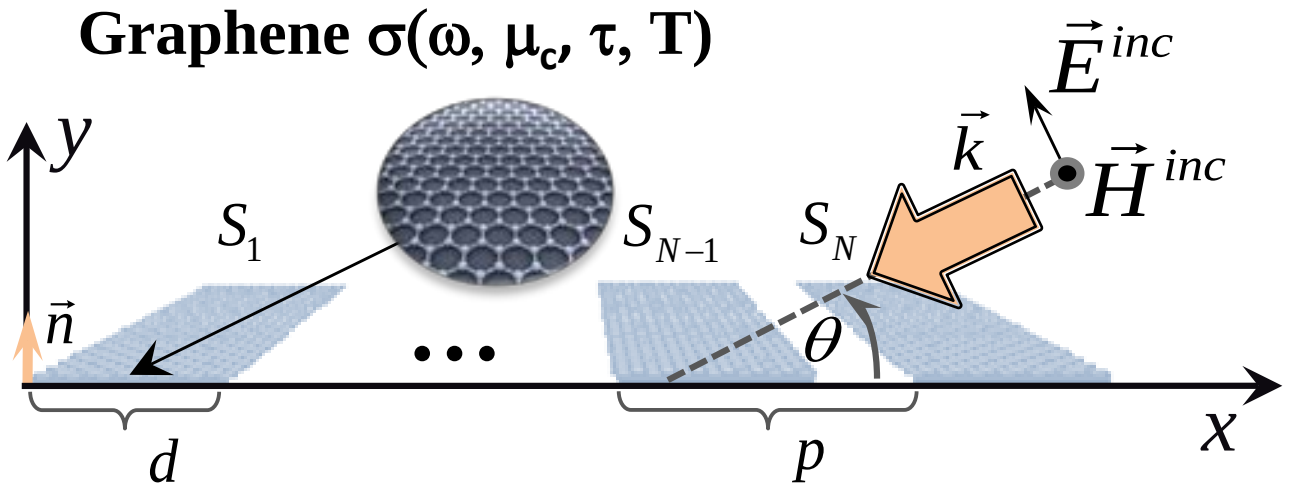

Fig. 1. Free-standing finite periodic grating of $N$ infinitely long (along the $z$-axis) graphene strips of the same width $d$ and period $p$.

Some of the THz applications of graphene include giant Faraday rotation [5], fixed and reconfigurable periodic frequency selective surfaces [10,11], plasmonic waveguides, switches and lenses [6,7] or absorbers and cloaks [8,11]. In addition, there has been much interest in patterned graphene materials to control their interaction with electromagnetic waves, that has led to novel antennas [9, 12].

Graphene strips have already attracted attention in the THz science community as attractive and easily manufactured components of plasmonic waveguides, antennas [12] and sensors. Single strip scattering has been studied in [13] and its wave-guiding properties in [14]. Plasmon-assisted resonances in the scattering and

This work was supported by the National Academy of Sciences of Ukraine via the State Target Program "Nanotechnologies and Nanomaterials," the European Science Foundation via the Research Networking Programme "Newfocus," the Swiss National Science Foundation via grant 133583, and the EU FP7 via Marie-Curie IEF grant "Marconi", with ref. 300966.

O. V. Shapoval and A. I. Nosich are with the Laboratory of Micro and Nano-Optics, Institute of Radio-Physics and Electronics of the National Academy of Sciences of Ukraine, Kharkiv 61085, Ukraine (e-mail: olga.v.shapoval@gmail.com, anosich@yahoo.com).

J. R. Mosig is with the Laboratory of Electromagnetics and Acoustics, Ecole Polytechnique Fédérale de Lausanne, CH-1015 Lausanne, Switzerland (e-mail:juan.mosig@epfl.ch)

J. Perruisseau-Carrier and, and J. S. Gomez-Diaz are with the group for Adaptive MicroNanoWave Systems, LEMA/Nanolab, École Polytechnique Fédérale de Lausanne, 1015 Lausanne, Switzerland (e-mail: julien.perruisseau-carrier@epfl.ch).

absorption by an infinite grating of coplanar graphene strips under normal incidence have been analyzed in [15] using the Fourier expansion method.

In this context, this paper proposes a rigorous study of the transmittance, absorption and scattering of THz waves by a coplanar graphene-strip grating. Compared with [15], we present a novel integral equation approach able to very efficiently analyze the problem under study. This allows us to greatly extend the results shown in [15], including two wave polarizations, different angles of the incoming waves, the presence of finite or an infinite number of strips, and different graphene parameters. In particular, we investigate the surface plasmon resonances, which depend on each strip conductivity and width, and the build-up of the Rayleigh anomalies at the wavelengths $\lambda = p(m \pm \sin\theta)^{-1}$, $m = 1, 2, ..$ caused by the grating periodicity and depending on the angle of incidence $\theta$ and the number of strips in the grating.

The very efficient analysis of the problems is achieved via the combined use of (i) integral equations (IEs) obtained using the surface-impedance (sometimes also called resistive-sheet) boundary conditions and (ii) Nystrom-type discretization that uses interpolation polynomials and appropriate quadrature formulas accounting for the kernel-function singularities and edge behavior. As known [1,2,10,16], graphene monolayer can be electromagnetically characterized with the aid of the following boundary conditions on a zero-thickness boundary:

$$(1/2)[\vec{E}_{tg}^{+} + \vec{E}_{tg}^{-}] = \sigma^{-1}\vec{n} \times [\vec{H}_{tg}^{+} - \vec{H}_{tg}^{-}], \qquad \vec{E}_{tg}^{+} = \vec{E}_{tg}^{-} \qquad (1)$$

imposed on the limiting values of the field components tangential (*tg*) to the top (+) and bottom (-) sides of layer.

Note that the jump in magnetic field is the electric surface current and the quantity $Z = 1/\sigma$ is called surface impedance. Conditions (1) were widely used some time ago in the analysis of the scattering by the gratings made of so-called resistive strips [17-19], with applications to microwave absorbers.

The Nystrom methods have already been used in the modeling of the wave scattering by perfectly electrically conducting strips [20-22], thin strips of conventional dielectrics [23] and finite periodic silver strip gratings in the optical band [24]. Their main merit is numerical efficiency and guaranteed convergence that entails easily controlled accuracy of computations. It allows rapid simulation of scatterers consisting of hundreds of micron-size graphene strips.

The paper is structured as follows. In Section II we present the scattering problem formulation, reduce it to a set of hyper-singular integral equations (IEs), and briefly explain the discretization scheme. Section III shows the actual rate of the algorithm convergence and validates the proposed approach using data from literature and commercially available full-wave software. Section IV presents the dependences of the scattering and absorption of THz waves by a finite and infinite graphene strip gratings on the frequency and geometrical parameters. Conclusions are summarized in Section IV. A brief review of graphene conductivity characterization and details of the numerical treatment are presented in Appendix. We assume that the electromagnetic field is time-harmonic $\sim e^{-i\omega t}$ and omit this dependence.

## II. Formulation, Integral Equations, and Nystrom-Type Algorithm

### *A. Formulation and Boundary Conditions*

Consider at first the two-dimensional (2-D) scattering and absorption of the H-polarized (vector $\vec{E}$ is across the strips) plane wave by a finite periodic grating made of $N$ identical coplanar graphene strips, in the THz frequency range. The corresponding freestanding geometry and the problem notations are shown in Fig. 1. The strips are assumed infinite along the *z*-axis, have zero thickness, width $d$ and period $p$. They are characterized with complex-valued graphene conductivity $\sigma(\omega, \mu_c, \Gamma, T)$ calculated via Kubo's formalism [16, 25] (see Appendix), where $\omega$ is the radian frequency, $\mu_c$ is the chemical potential, $\Gamma = (2\tau)^{-1}$ is the phenomenological scattering rate that is assumed to be independent of energy ($\tau$ is the relaxation time of charge carriers), and $T$ is the temperature [1,2,11].

As known, in a 2-D scattering problem one has to find a scalar function $H_z^{sc}(\vec{r})$ that is the scattered magnetic field *z*-component [23]. Here, the total field function $H_z(\vec{r}) = H_z^{sc}(\vec{r}) + e^{-ik(x\cos\theta + y\sin\theta)}$, $\vec{r} = (x, y)$ must satisfy the Helmholtz equation off the strip surface $S = \cup_{j=1}^{N} S_j$, where $S_j = \{(x,0) : x \in [a_j, b_j]\}$, and $a_j$ and $b_j$ are the $j$ strips endpoints. The conditions (1) at $\vec{r} \in S$ take the form as

$$\partial\vec{H}_z(\vec{r}) / \partial\vec{n} = -ik(\sigma Z_0)^{-1}[H_z^{+}(\vec{r}) - \vec{H}_z^{-}(\vec{r})], \qquad (2)$$

where $k = \omega / c$ is the free-space wavenumber ($c$ is the space velocity) and $Z_0 = (\mu_0 / \varepsilon_0)^{1/2}$ is the free-space impedance.

For solution uniqueness, the formulation of the problem must be completed with the local integrability of power (edge condition) and the radiation condition at infinity [23,24].

### B. Hyper-Singular Integral Equations

To satisfy Helmholtz equation and radiation condition, we seek the scattered field as a sum of double-layer potentials,

$$H_z^{sc}(\vec{r}) = \sum_{j=1}^{N} \int_{S_j} w_j(\vec{r}\,') \frac{\partial G(\vec{r},\vec{r}\,')}{\partial \vec{r}\,'} d\vec{r}\,', \tag{3}$$

where $G(\vec{r},\vec{r}\,') = (i/4) H_0^{(1)}(k|\vec{r}-\vec{r}\,'|)$ is the Green function. Note that the unknown functions are electric currents induced on the strips, $w_j(\vec{r}) = H_z^+(\vec{r}) - \vec{H}_z^-(\vec{r})$, $\vec{r} \in S_j$, $j = 1,...,N$.

Using condition (2) and the properties of the limit values of the double-layer potentials, we obtain the following set of $N$ coupled IEs for $w_j(x^j)$ where $x^j \in [a_j, b_j]$,

$$4(\sigma Z_0)^{-1} w_i(x_0^i) + \sum_{j=1}^{N} \int_{a_j}^{b_j} w_j(x^j) \frac{H_1^{(1)}(k|x^j - x_0^i|)}{|x^j - x_0^i|} dx^j = f(x_0^i), \tag{4}$$

where $f(x_0^i) = 4\sin\theta e^{-ikx_0^i \cos\theta}$, $x_0^i \in S_i$ $(i = 1,...,N)$. Note that these IEs are fully equivalent to the original boundary-value problem. The integrals in (4) are understood in the sense of finite part of Hadamard, and unknowns can be represented as products $w_j(x) = \tilde{w}_j(x)[x-(j-1)p]^{1/2}[(j-1)p+d-x]^{1/2}$, where $\tilde{w}_j(x)$ is smooth and non-singular at the $S_j$ endpoints.

### C. Nystrom-Type Discretization

It is easy to see that the kernel functions in (5) display both hyper-type and logarithmic singularities, if $x \to x_0$, because

$$\frac{H_1^{(1)}(k|x-x_0|)}{|x-x_0|} \sim ik \frac{\ln|x-x_0|}{\pi} - \frac{2}{\pi k |x-x_0|^2} \tag{5}$$

In fact, it is this hyper-singularity, together with the step-like dependence of conductivity $\sigma$ on $x$ in the grating plane ($\sigma = 0$ out of strips), which makes the projection of (4) on the Floquet basis functions (this is Fourier-expansion method) a divergent algorithm. In contrast, we discretize the IEs (4) using Nystrom-type method based on the Gauss-Chebyshev quadrature formulas of interpolation type of the $n$-th order (with Chebyshev weight). As the discretization and collocations nodes, we choose the nulls of the Chebyshev polynomials of the second kind. This leads to $N \times N$ block-type matrix equation with $n \times n$-sized blocks. After solving it numerically we obtain approximate solution of IEs in the form of $N$ interpolations polynomials for the unknown surface currents. Then the field (3) is easily reconstructed in the near and far zone of the strip grating. Our meshless numerical algorithm is efficient and reliable and has theoretically guaranteed convergence (at least as $1/n$ as follows from the convergence of quadratures) and controlled accuracy of computations. For more mathematical details about the Nystrom-type discretizations see [20-24].

## III. Nystrom-Type Algorithm Convergence and Method Validation

### A. Scattering and Absorption Characteristics

To study the features of the plane wave scattering and absorption by a *finite* grating of $N$ graphene strips, the figures-of-merit are not the reflectance, transmittance and absorbance usually introduced in infinite-grating scattering problems, but the total scattering cross section (TSCS) and the absorption cross-section (ACS) [26], p.13. Besides, the monostatic radar cross-section (RCS) is known to characterize the power that is reflected back to the source [27]. As the scattered field in the far zone ($r \to \infty$) takes the form $H_z^{sc} \sim (2/i\pi kr)^{1/2} e^{ikr} \Phi(\varphi)$, where the radiation pattern is found as

$$\Phi(\varphi) = (k\sin\varphi/4) \sum_{j=1}^{N} \int_{S_j} w_j(x) e^{-ikx\cos\varphi} dx, \tag{6}$$

the following expressions are obtained for TSCS and RCS:

$$\sigma_{tsc} = (2/\pi k) \int_0^{2\pi} |\Phi(\varphi)|^2 d\varphi, \quad \sigma_{rsc} = 4|\Phi(\theta)|^2 / k \tag{7}$$

The ACS is obtained from the near-field integration,

$$\sigma_{abs} = \mathrm{Re}(\sigma Z_0)^{-1} \sum_{j=1}^{N} \int_{S_j} |w_j(x)|^2 dx. \tag{8}$$

and the optical theorem (power conservation law) states that $\sigma_{tsc} + \sigma_{abs} = -(4/k)\,\mathrm{Re}\,\Phi(\varphi + \theta)$.

### B. Algorithm Convergence

As convergence is guaranteed by the general theorems, the accuracy can be, in principle, at the level of machine precision. For the demonstration of the actual rate of convergence, we have computed the root-mean-square deviations $e_w(n) = |\theta_w^n / \theta_w^{2n} - 1|$ of the uniform norms of the surface current functions, $\theta_w^2 = \sum_{j=1}^{N} \int_{-1}^{1} |w_j(t)|^2 \, dt$, versus the order of discretization polynomial, $n$. The results are shown in Fig. 2.

As visible, the proposed Nystrom method ensures algebraic convergence of the approximate solution to the accurate one with increasing discretization order $n$.

For instance, to achieve 4-digit accuracy in the near field analysis one can take $n$ = 55 in a frequency range up to 10 THz with a strip width of 20 μm.

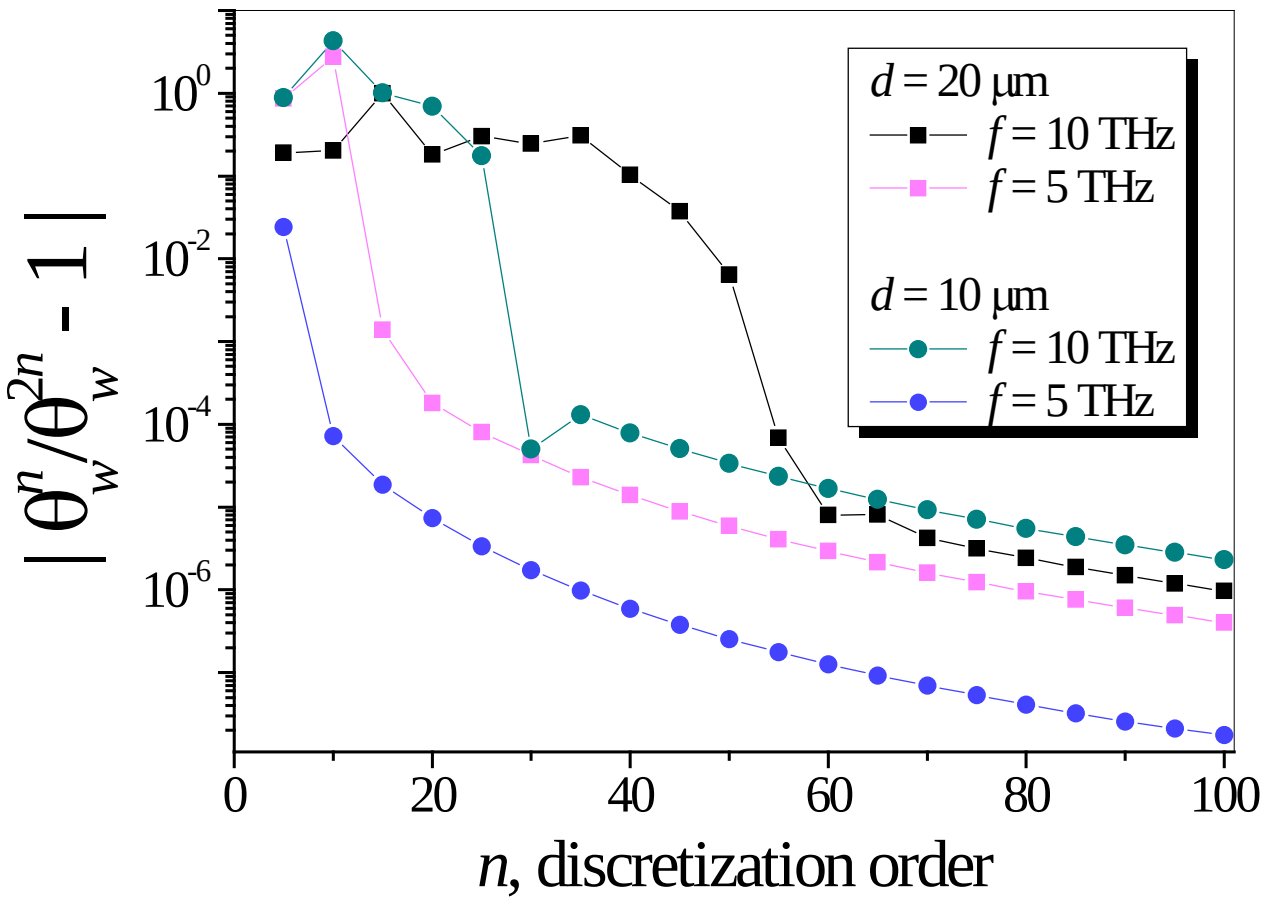

Fig. 2. Computation error $e_w(n)$ as a function of the discretization order $n$ for a stand-alone graphene strip of the width $d$ = 10 and 20 μm at $f$ = 5 and 10 THz under the normal incidence of a H-wave; graphene parameters are τ = 1 ps, $\mu_c$ = 0.13 eV and $T$ = 300 K.

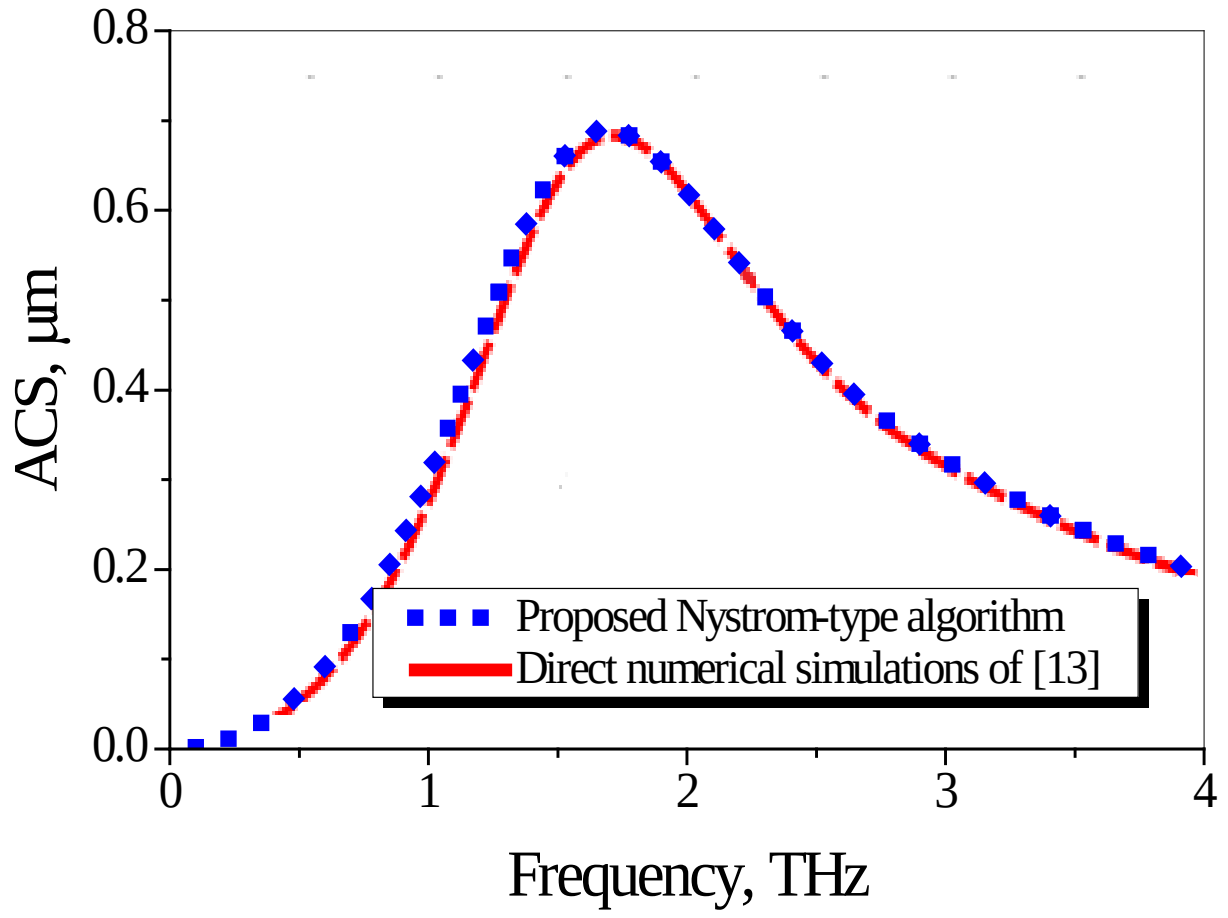

Fig. 3. ACS versus the frequency in THz range for a normally incident H-wave scattered by a stand-alone graphene strip of $d$ = 5 μm calculated using proposed Nystrom-type algorithm (blue curve) and numerical simulations of [13] (red curve); graphene parameters are τ = 0.1 ps, $\mu_c$ = 0.0 eV and T = 300 K.

## C. *Algorithm Validation*

The ACS results obtained using the proposed Nystrom-type algorithm provides very good agreement with the data of [13] for a single infinitely long graphene strip of $d$ = 5 μm suspended in free space (see Fig. 3).

A comparison with the transmittance and absorbance of an infinite graphene-strip grating computed using the Fourier–Floquet expansion method of [15] has shown good qualitative agreement, especially in the location of resonances. However, it is known that this approach may suffer of the lack of convergence – see discussion in Section II C.

Besides, we have compared our results for a stand-alone graphene strip with the data generated by the commercial software HFSS. In Fig. 4, we present the graphs of the monostatic RCS versus frequency for a strip of $d$ = 20 μm, $\tau$ = 1 ps and $\mu_c$ = 0.2 eV under normal incidence $(\theta = \pi/2)$ calculated with the Nystrom-type algorithm (blue curve) and HFSS (red curve). HFSS results agree well with the proposed method for low frequencies; however this is not so above 2.2 THz where the mesh becomes insufficiently dense. Making it denser leads to prohibitively large simulation times. Overall obtaining the presented curve with HFSS requires about 9 hours of computation time, while modeling a similar problem using proposed Nystrom-type algorithm takes only 40 seconds with a $10^{-3}$ THz step, in the wider frequency range.

It should be noted that computing the TSCS or ACS with HFSS would lead to completely unrealistic simulation times, and therefore we provide only the RCS comparison.

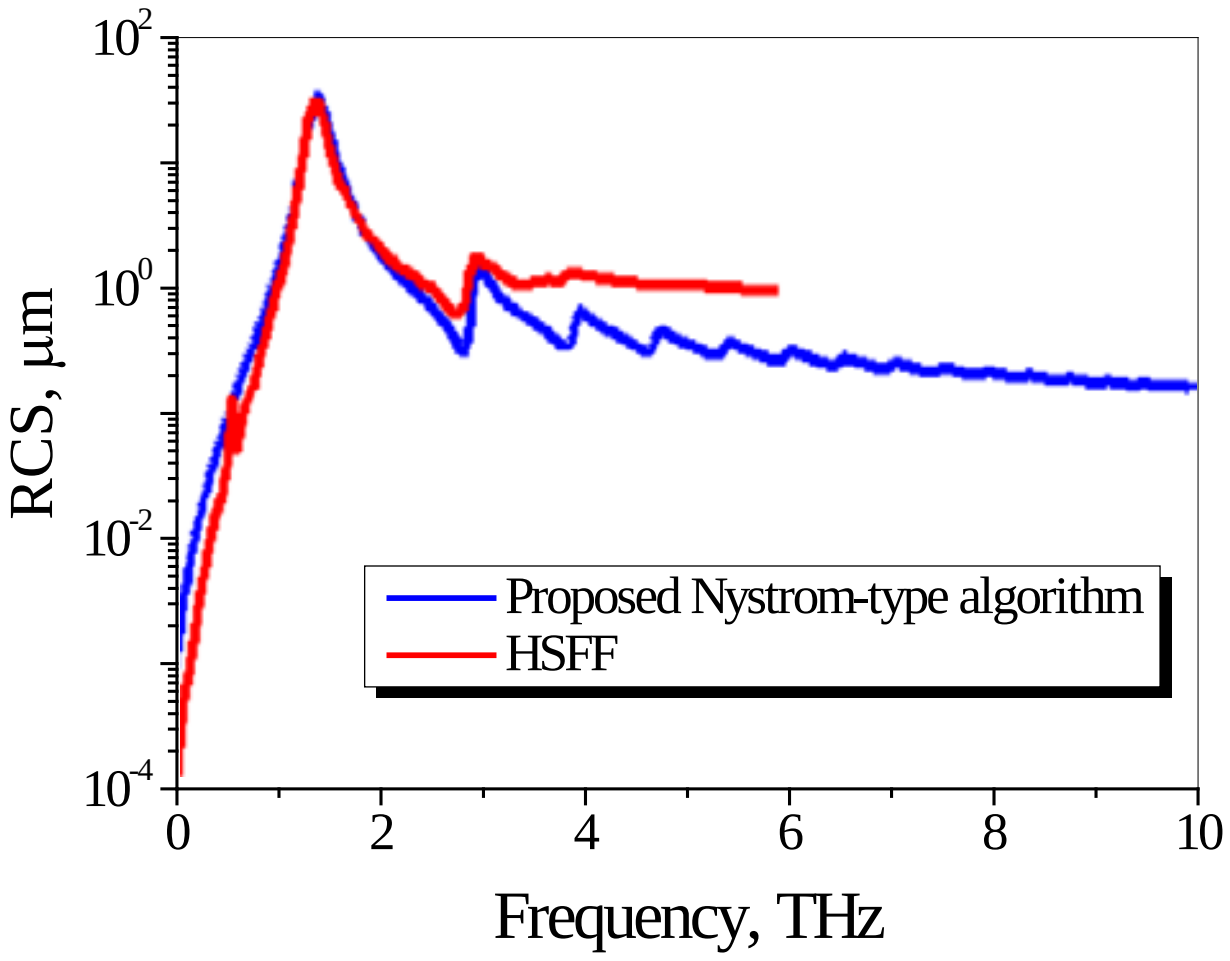

Fig. 4. RCS versus the frequency for a normally incident H-wave scattered by a stand-alone graphene strip of $d$ = 20 μm calculated using Nystrom-type algorithm (blue curve) and HFSS (red curve); graphene parameters are τ = 1 ps, $\mu_c$ = 0.2 eV and T = 300 K.

## IV. Numerical Results and Discussion

### A. *Stand-Alone Microsize Graphene Strip*

Stand-alone microsize graphene strip illuminated by the H-polarized wave in the THz frequency range demonstrates a variety of surface plasmon resonances.
In Fig. 5 (a) and (b) we show the TSCS and ACS as a function of frequency for several strips of different widths $d$ and two incidence angles, $\theta = \pi/2$ (solid curves) and $\theta = \pi/4$ (dotted curves). Here, the chemical potential is $\mu_c$ = 0.13 eV, $\tau = 1\,\text{ps}$ and the room temperature ($T$ = 300 K) is assumed. As can be observed, the TSCS and ACS spectra strongly depend on the strip width. The wider strips demonstrate larger numbers of resonances on the localized surface plasmons, shifting-down in frequencies. We also observe little dependence of these parameters on the incident angle of the incoming waves.

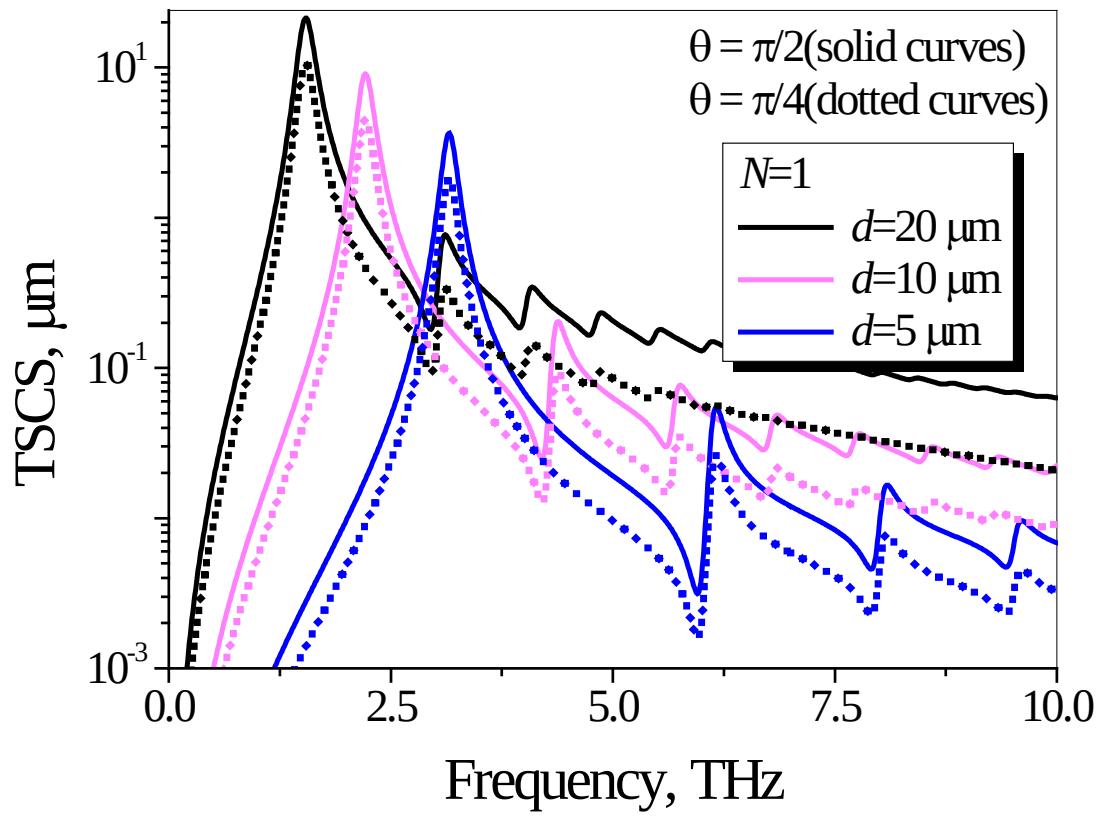

(a)

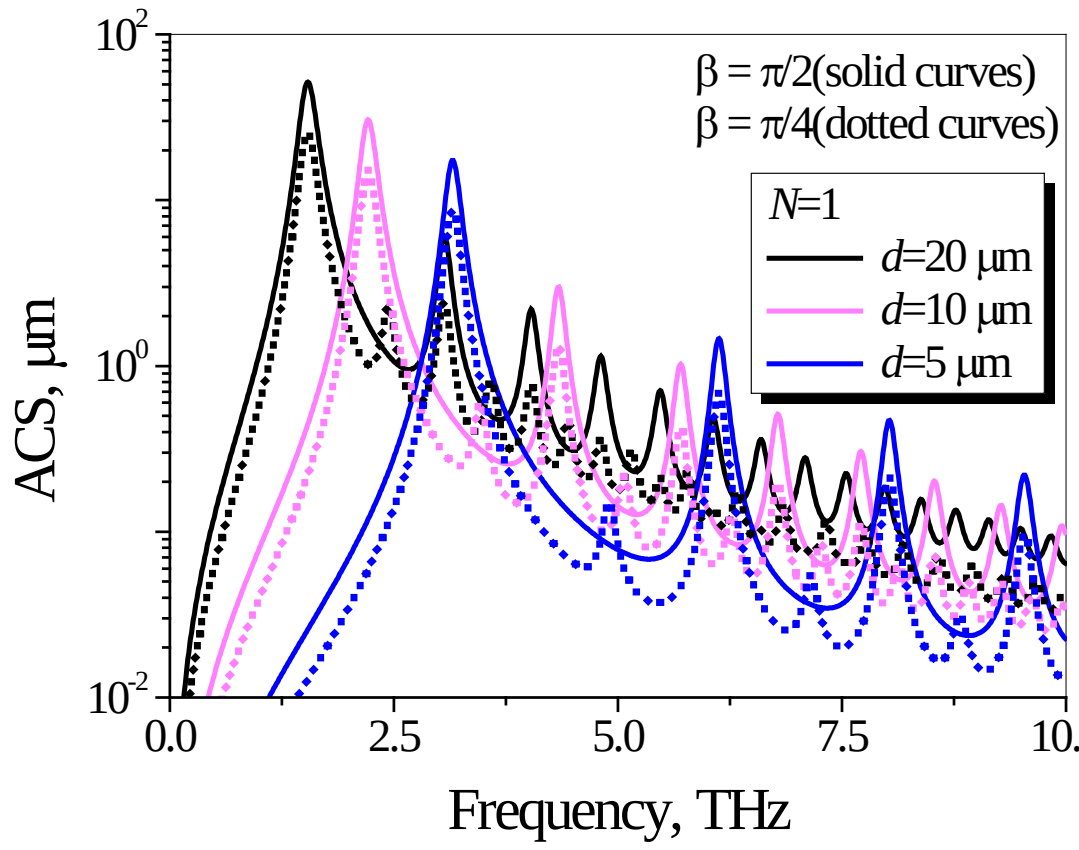

(b)

Fig. 5. TSCS (a) and ACS (b) versus the frequency for the normally (solid curves) and inclined (dotted curves) incident H-wave scattering by a stand-alone graphene strip of varying width $d$; graphene parameters are τ = 1 ps, $\mu_c$ = 0.13 eV and T = 300 K.

Figs. 6 (a)-(b) display the near-field patterns of $|H_z|$ at the first four surface-plasmon resonances $H_m$ for the scattering by a graphene strip of $d$ = 20 μm under the normal $\theta = \pi/2$ (a) and inclined $\theta = \pi/4$ (b) incidence, respectively. Here only odd-index resonances are excited at the normal incidence because of their symmetry across the $y$-axis, and both odd and even ones appear under the inclined incidence.

It should be noted that each $H_m$ resonance is formed as a Fabry-Perot like standing wave because of the reflections of the surface-plasmon natural wave of a graphene layer from the strip edges. The complex-valued propagation constant of this wave can be obtained analytically from (1),

$$\beta_{pl} = k[1-(\sigma Z_0/2)^2]^{1/2}, \tag{9}$$

and then the associated resonance frequencies satisfy

$$\operatorname{Re}\beta_{pl} d \approx m\pi, \quad m = 1, 2, ... \tag{10}$$

We have also analyzed the spectral response of a free-standing graphene strip for different values of chemical potential $\mu_c$ and relaxation time $\tau$.

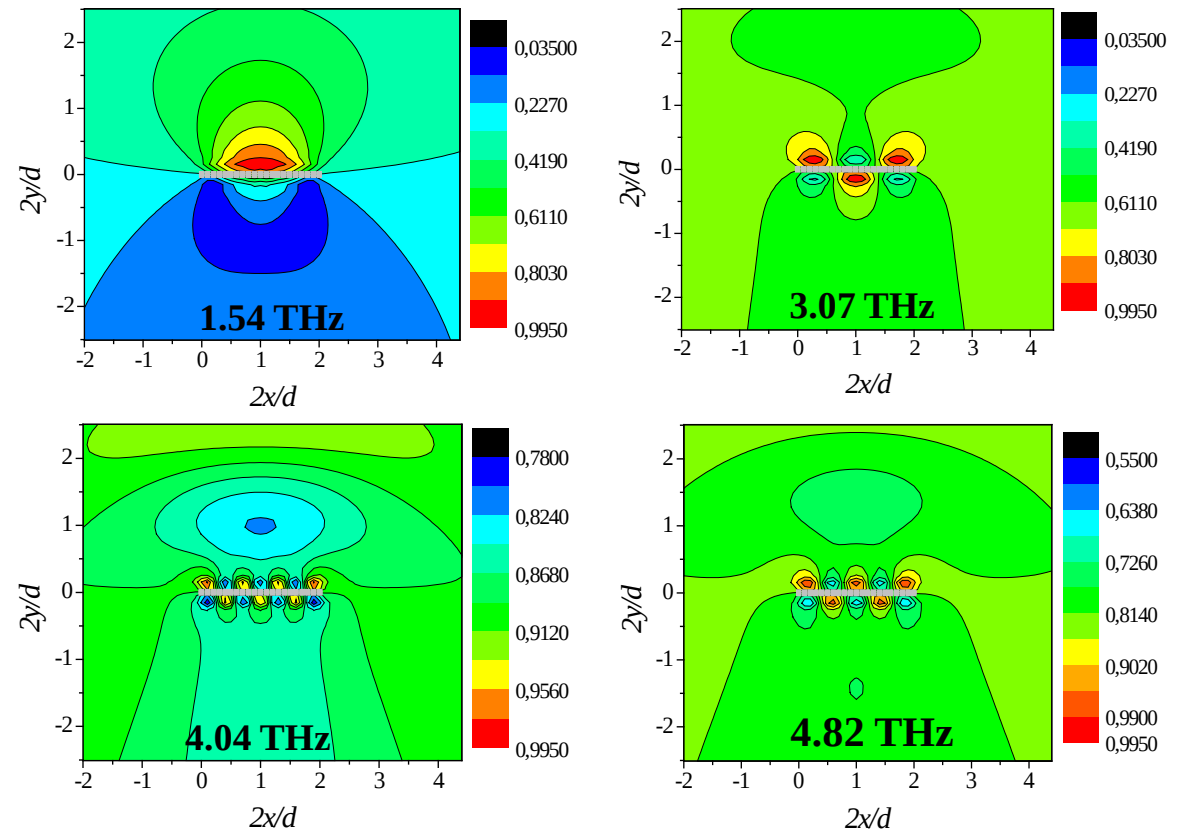

(a)

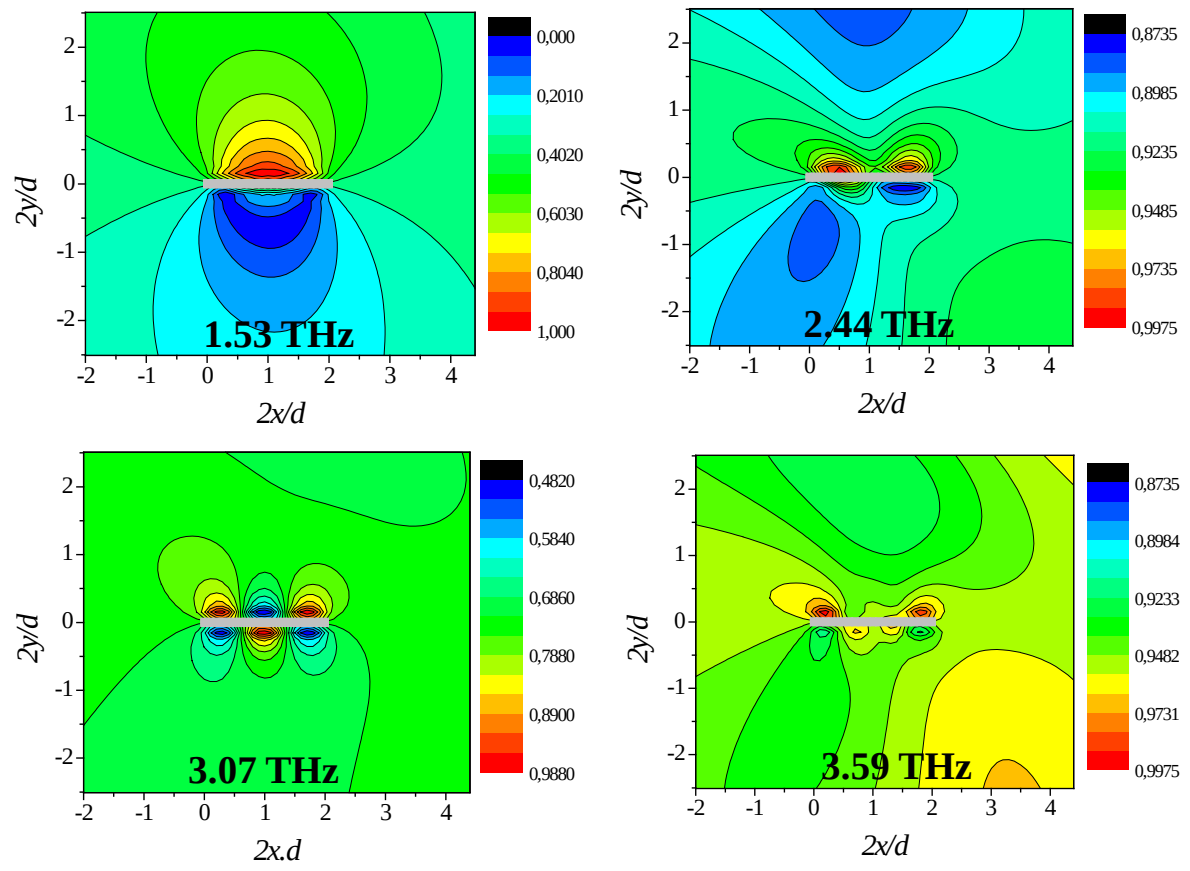

(b)

Fig. 6. Near-field patterns for a stand-alone graphene strip under the normal $\theta = \pi/2$ (a) and inclined $\theta = \pi/4$ incidence (b) for the four lower resonant frequencies.

Fig. 7 shows that the magnitudes of the surface-plasmon resonances are quite sensitive to the relaxation time changes. The peaks of TSCS and ACS became more pronounced if $\tau$ increases due to the smaller dissipation losses (see Appendix). Unlike ACS, all resonances in TSCS except of $H_1$ demonstrate asymmetric Fano shapes.

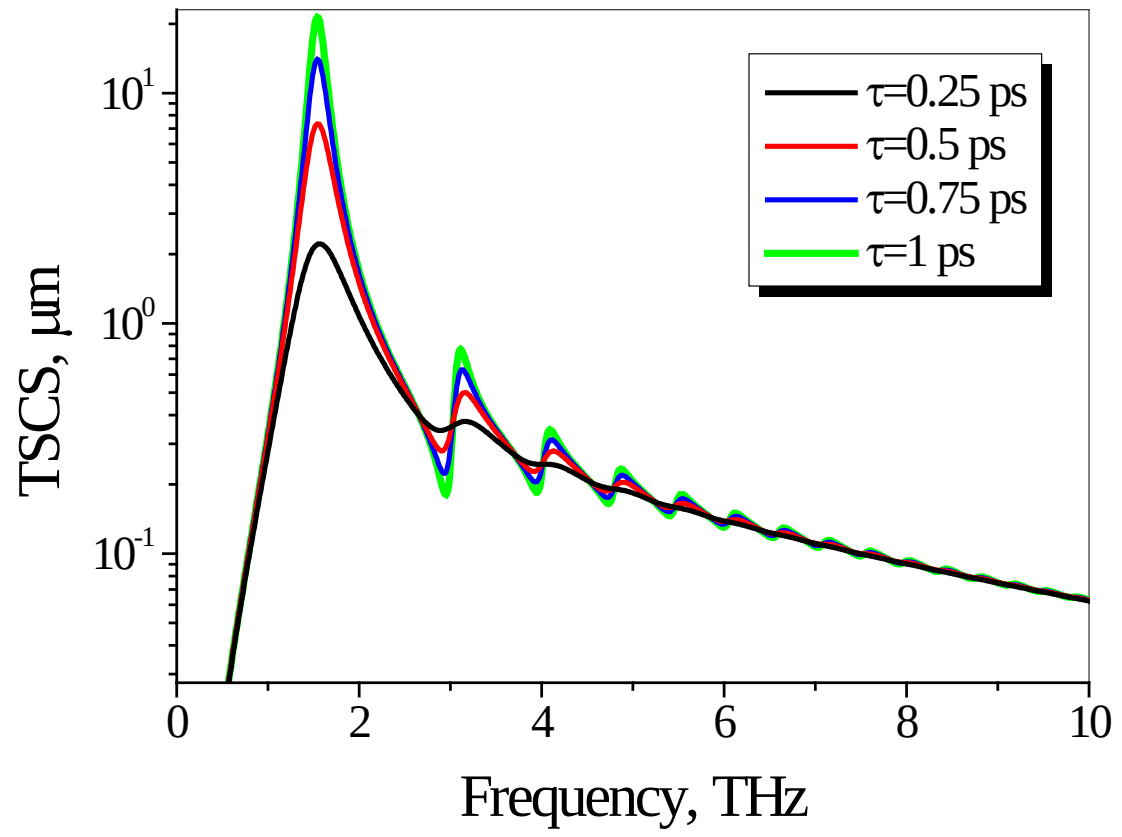

(a)

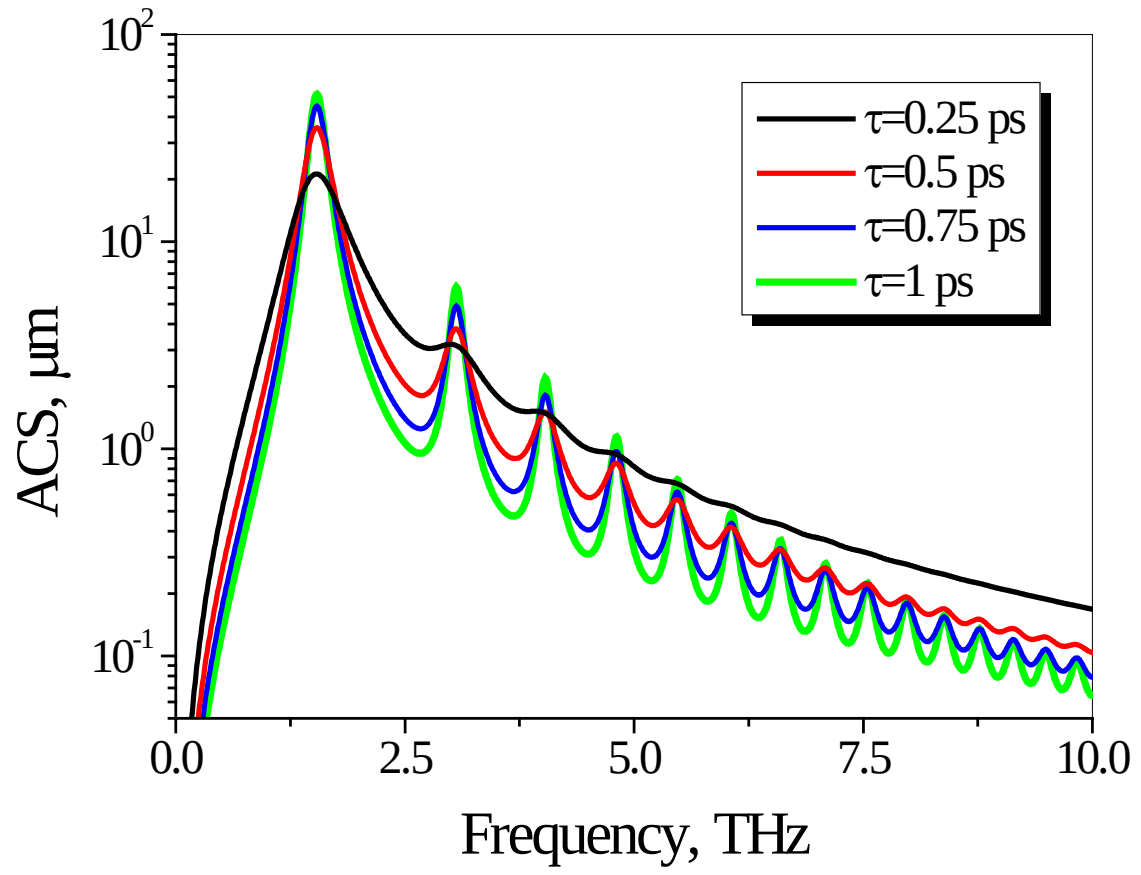

(b)

Fig. 7. Same study as in Fig. 5 but for a normally incident H-wave and a strip width $d$ = 20 μm for different values of relaxation time, $\tau$ = 0.25, 0.5, 0.75 and 1 ps and fixed chemical potential $\mu_c$ = 0.13 eV.

In turn, Fig. 8 shows that an increase of $\mu_c$ up-shifts the resonance frequencies and decreases losses in graphene. This is in agreement with the surface impedance of graphene (see Appendix) and explains the behavior of TCS and ACS in Fig. 8, where significant enhancement of the absorption coefficient is obtained with a slight change in the chemical potential $\mu_c$.

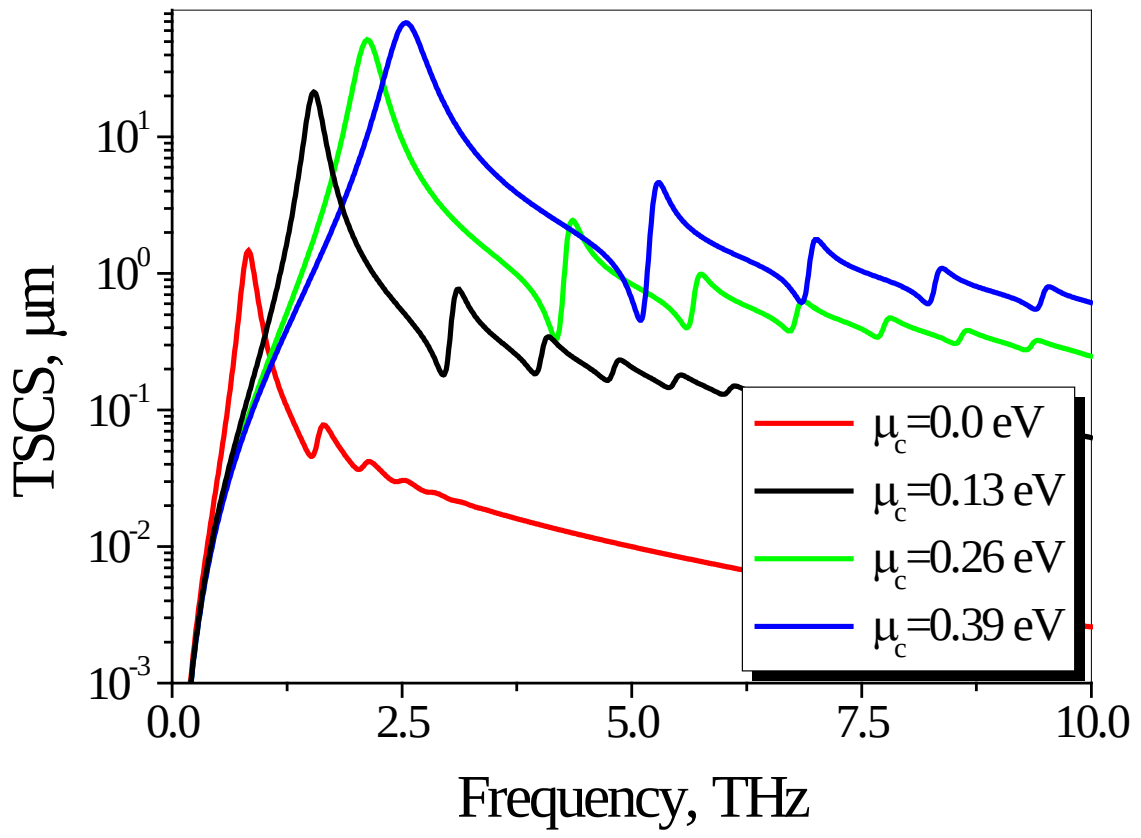

(a)

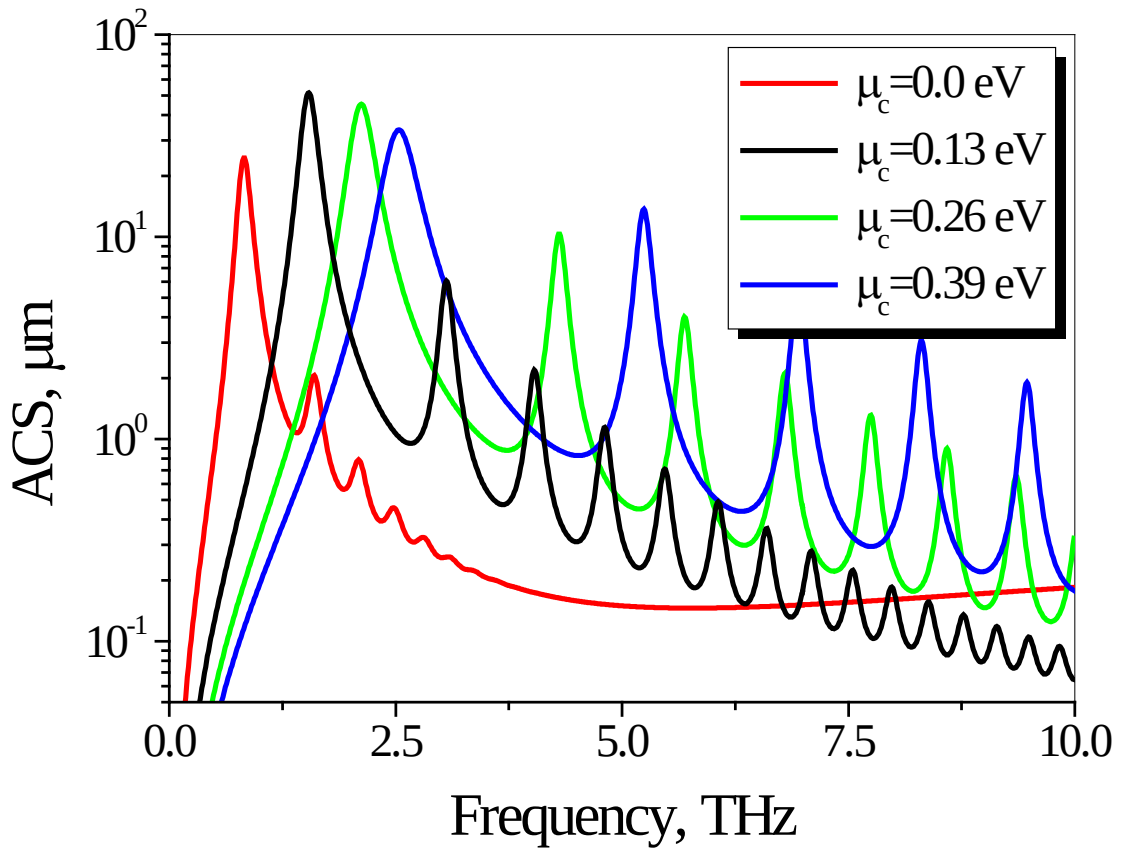

(b)

Fig. 8. Same study as in Fig. 7 but for a fixed relaxation time of $\tau$ = 1 ps and different values of the chemical potential $\mu_c$ = 0.0, 0.13, 0.26 and 0.39 eV, $T$ = 300 K.

### *B. Resonance Prediction According to the Strip Width*

In order to obtain a better insight into single-strip resonances, Fig. 9 presents the dependences of the resonance frequencies of the first to the fourth-order plasmon resonances versus the strip width. For comparison, the stars show the same values obtained by solving the approximate equation (10).

This result demonstrates that a wider free-standing graphene strip present more resonances in the considered THz range, with the lowest of them down-shifted in frequencies.

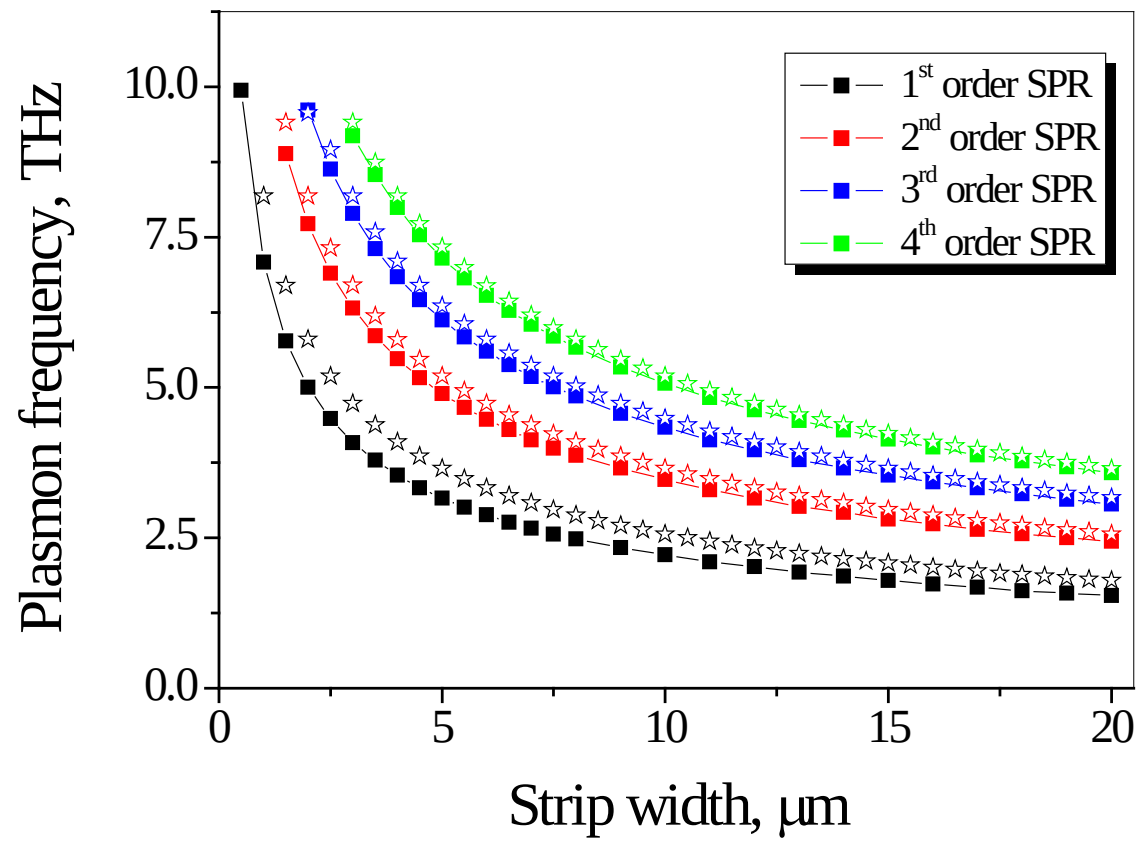

Fig. 9. Dependence of the 1st, 2nd, 3rd and 4th order plasmon resonance frequencies versus graphene strip width; graphene parameters are T = 300 K, τ = 1 ps and $\mu_c$ = 0.13 eV.

### *C. Gap Size Effects in Finite Graphene Strip Gratings*

In Fig. 10, we present the plots of TSCS (a) and ACS (b) of the grating of three graphene strips of widths $d$ = 20 μm, focusing on the influence of the gap size, $g$.

These results show that, if the gaps between the strips are large, the resonances keep their shapes and positions and the scattering and absorption spectra per one strip show the same behavior as for a stand-alone strip. Narrower gaps provide, however, a stronger wave coupling between the strips so that the whole terahertz-range spectral response of the three 20-μm strip configuration gradually transforms into the response of one 60-μm strip.

In Fig. 11, we consider a larger grating of $N$ = 10 strips of widths $d$ varying from 10 to 60 μm and fixed period $p$ = 70 μm. As far as the airgaps get smaller, the interaction between the strips becomes larger and the plasmon resonances shift down in frequency.

### *D. Comparison between Finite and Infinite Gratings for H- and E-wave Scattering*

It is interesting to compare the THz properties of infinite and finite gratings made of graphene strips. However it is not obvious how to select a common figure-of-merit. We have found that the reflectance of a plane wave by a finite grating can be introduced as the part of TSCS associated with the power scattered into the upper half-space. The transmittance of finite grating can be derived in similar manner however with account of the optical theorem. In addition, we normalize these values by the strip electric width $kd$, the number of strips $N$ and $\zeta = d/p$, that leads us to expressions

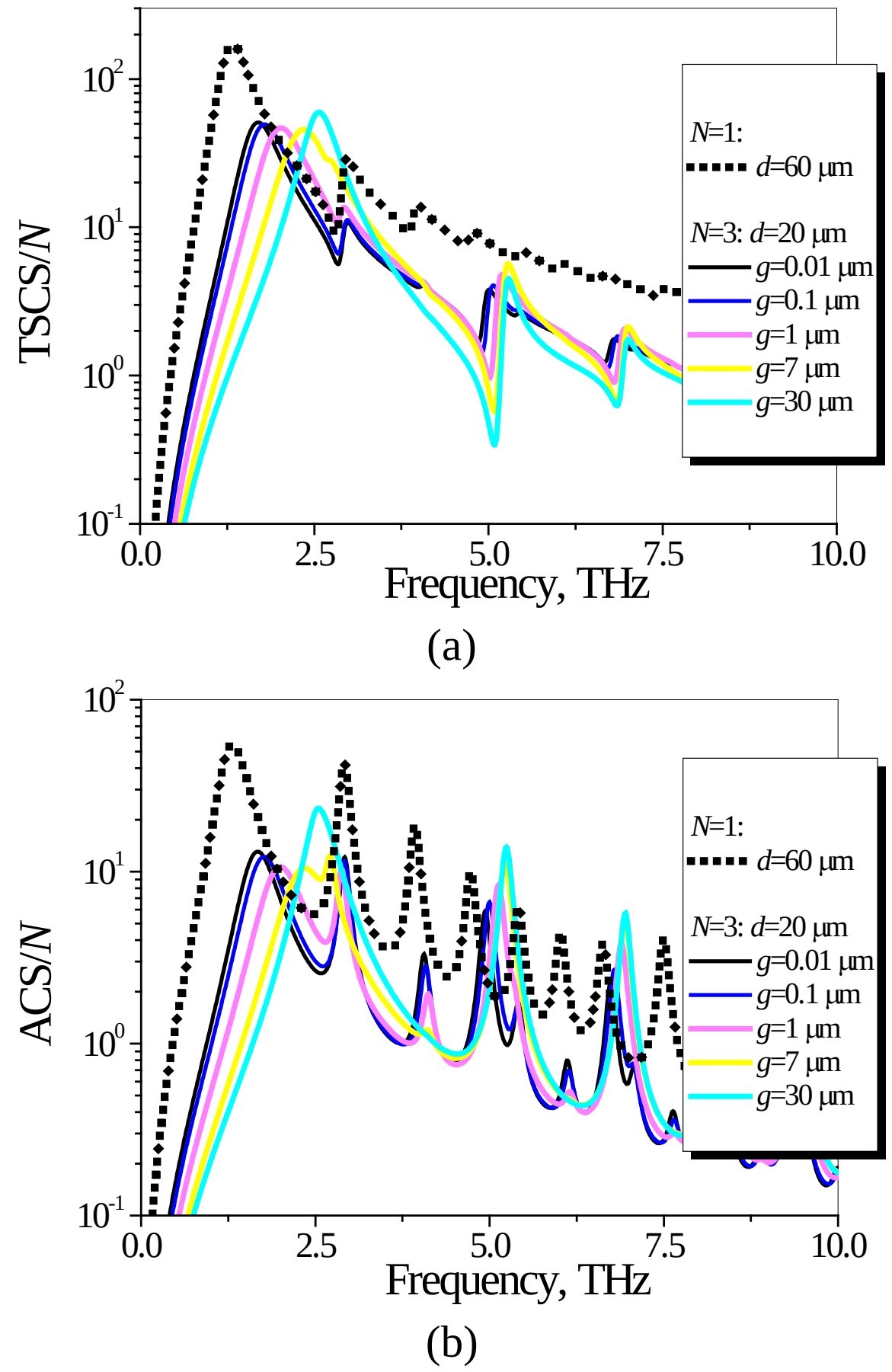

Fig. 10. Normalized TSCS (a) and ACS (b) versus the frequency for a normally incident H-wave scattered by a grating of $N$ = 3 strips of width $d$ = 20 μm for the different gaps between strips $g$; graphene parameters are T = 300 K, $\tau$ = 1 ps and $\mu_c$ = 0.39 eV.

$$R_{fin} = 2\zeta / (\pi N k d) \int_0^{\pi} |\Phi(\varphi)|^2 \, d\varphi, \tag{11}$$

$$T_{fin} = 1 + \frac{2\zeta}{\pi N k d} \int_{\pi}^{2\pi} |\Phi(\varphi)|^2 \, d\varphi + \frac{4\zeta}{N k d} \operatorname{Re} \Phi(\varphi + \theta), \tag{12}$$

while the absorbance can be found as $A_{fin} = 1 - (T_{fin} + R_{fin})$ that follows from the conservation of power.

These quantities can be conveniently compared to the reflectance, transmittance and absorbance of infinite grating.

In Fig. 12, we present such comparison for the H- and E-wave scattering by finite ($N$ = 10 and 50) and infinite strip gratings with $d$ = 20 μm and $p$ = 70 μm, at the normal incidence. Note that the alternative case of the E-polarized wave scattering (vector $\vec{E}$ is parallel to the strips) reduces to a set of $N$ coupled IEs with logarithmic singularities in the kernels. They are solved similarly to section II using Nystrom-type discretization with Gauss-Legendre quadrature formulas [21, 23]. This is because the currents tend to finite values at the edges of resistive

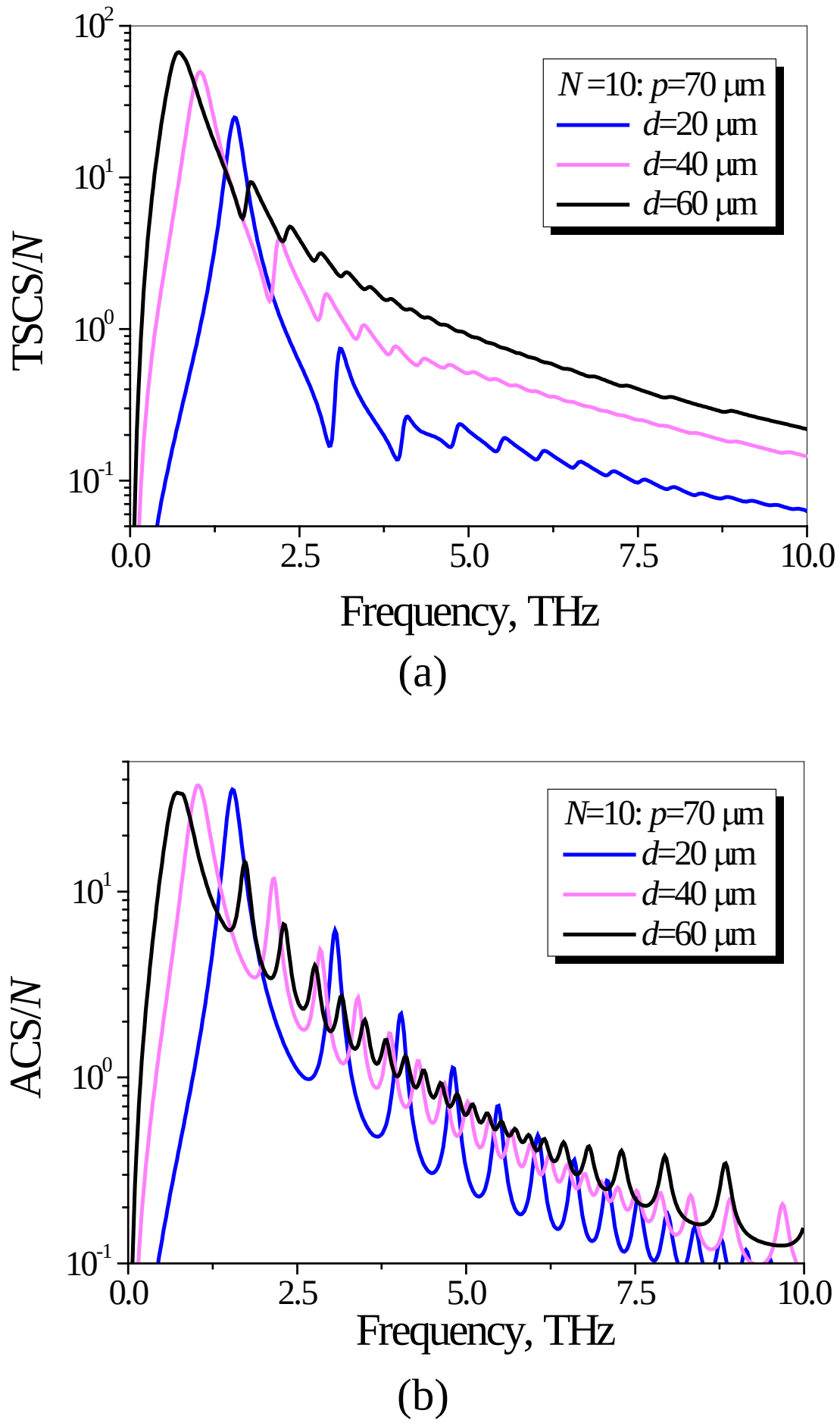

Fig. 11. Same study as in Fig. 7 but for a normally incident H-wave scattered by a grating of $N$ = 10 strips with different values of width $d$ and fixed period $p$ = 70 μm; graphene parameters are $\tau$ = 1 ps, $\mu_c$ = 0.13 eV and T = 300 K.

strips and hence no weight is needed

In the case of infinite strip grating, we have used the same algorithm as explained in Section II where Green's function of the free space is replaced with the pseudo-periodic Green's function (having the phase factor of $\exp(ikp\cos\theta)$ ). As both functions have the same logarithmic singularity, this does not lead to new quadrature formulas.

As one can see, the H-polarization case demonstrates plasmon resonances in the THz range and a gradual build-up of the Rayleigh anomalies at the associated wavelengths $\lambda = p/m$ , $m = 1,2$ . As expected, in the E-polarization case no plasmon resonances appear, however more pronounced Rayleigh anomalies are build and show enhanced transmission: one where the wavelength equals to period, $f$ = 4.283 THz and the other where it makes a half of period, $f$ = 8.565 THz.

As visible, even 10 strips provide normalized reflectance and absorbance values very close to the infinite grating values in the whole band of frequencies from 0.1 to 10 THz, except for the narrow bands around the Rayleigh anomalies. Interestingly, either of polarizations does not display high-quality grating resonances that have been recently reported for the silver nanostrip gratings in the visible range [28] in the H-polarization case.

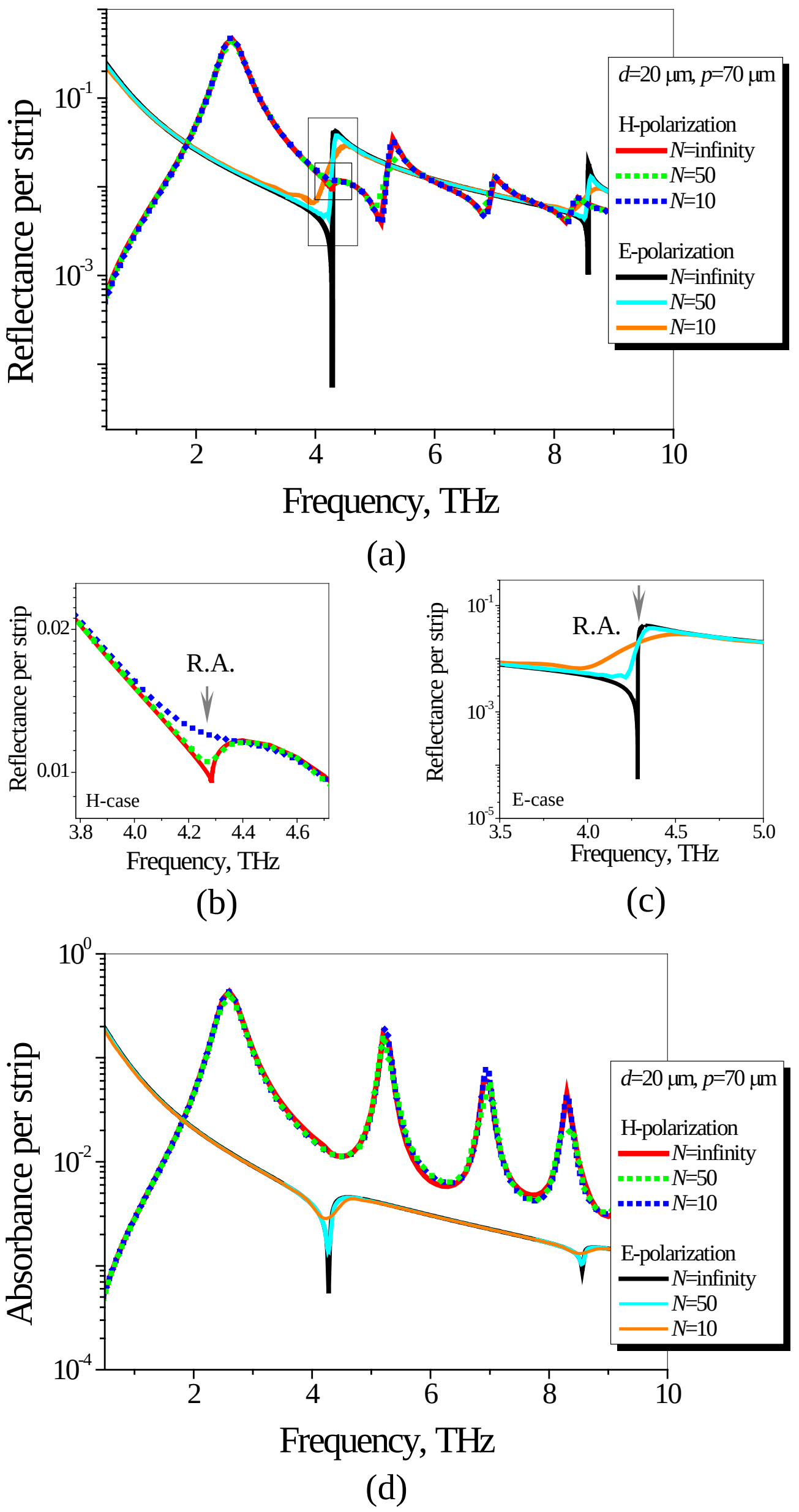

Fig. 12. Reflectance (a) and absorbance (d) per one strip for the normally incident plane waves of two polarizations scattered by the infinite and $N$-strip gratings of $d$ = 20 μm and $p$ = 70 μm. Panels (b) and (c) show zoomed reflectance near the Rayleigh anomaly wavelength in H-case and E-case, respectively; graphene parameters are $\tau$ = 1 ps, $\mu_c$ = 0.39 eV and $T$ = 300 K.

Fig. 13 shows the near E-field in the vicinity of the Rayleigh anomaly at $f$ = 4.274 μm for a $N$ = 50 strip grating (around the strips 25 to 28) with width $d$ = 20 μm and period $p$ = 70 μm, demonstrating good transparency. The corresponding near H-field patterns for the H-wave scattering at the first-order plasmon resonance $f$ = 2.6 THz (b) and at the first Rayleigh anomaly frequency $f$ = 4.27 THz (c) demonstrate high reflectivity of the incident plane wave and slightly increased transparency, respectively.

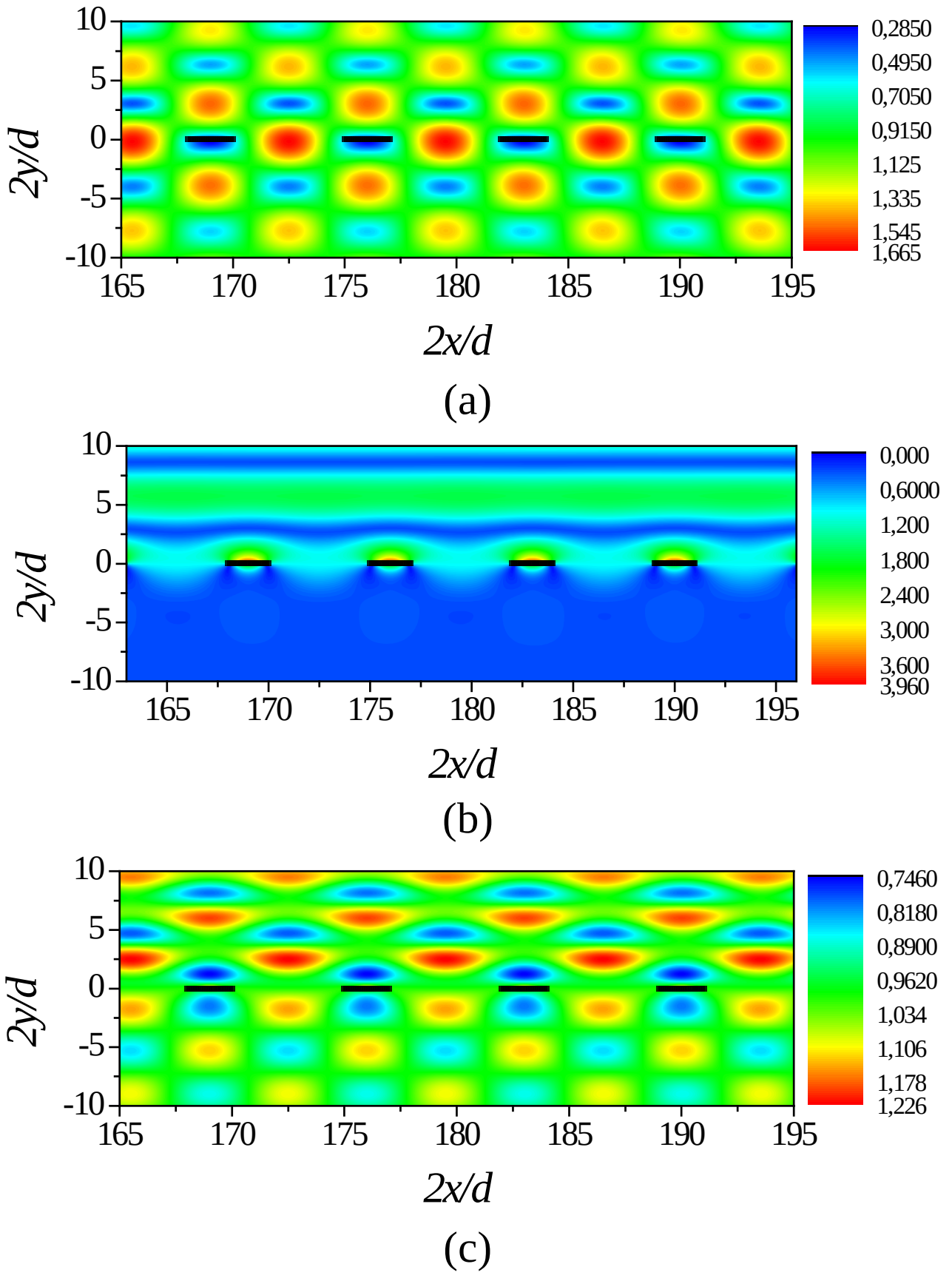

Fig.13. Near E-field pattern around the strips 25 to 28 for a E-wave normally incident at the grating of $N$ = 50 graphene strips of $d$ = 20 μm and $p$ = 70 μm near Rayleigh anomaly frequency $f$ = 4.274 THz (a); H-field patterns for the H-wave incidence in the plasmon resonance at $f$ = 2.6 THz (b) and at Rayleigh anomaly (c).

## V. Conclusions

In summary, we have presented a numerically efficient and accurate analysis of the scattering and absorption of plane waves by finite and infinite coplanar graphene-strip gratings in the THz frequency range. Unlike standard full-wave commercial software, the developed meshless algorithm is based on the singular IEs and Nystrom-type discretizations, which provides a theoretically guaranteed convergence up to machine precision.

Application of the method shows the presence of surface plasmon resonances in the THz range in the case of H-polarization, and a gradual build-up of the Rayleigh anomalies as the strip number gets larger in the both polarization cases. We have also investigated the tunability of the mentioned resonances with respect to the graphene chemical potential, relaxation time and the grating configuration.

These effects can be potentially exploited in the design of tunable THz range filters, frequency selective surfaces, and ultrathin absorbing panels for electromagnetic compatibility.

## Acknowledgment

We are grateful to A.Y. Nikitin and L. Martin-Moreno for valuable discussions.

## Appendix: Graphene Conductivity

Graphene conductivity is characterized applying the Kubo formula [1,2,11,16],

$$\sigma(\omega,\mu_c,\Gamma,T)=\xi[(\omega+i2\Gamma)^{-2}\int_0^{\infty}\varepsilon\left(\frac{\partial f_d(\varepsilon)}{\partial\varepsilon}-\frac{\partial f_d(-\varepsilon)}{\partial\varepsilon}\right)\partial\varepsilon$$
$$-\int_0^{\infty}\frac{f_d(-\varepsilon)-f_d(\varepsilon)}{(\omega+i2\Gamma)^2-4(\varepsilon/\hbar)^2}\partial\varepsilon], \qquad \text{(A1)}$$

where $\xi=-iq_e^2(\omega+i2T)(\pi\hbar^2)^{-1}$, and $\varepsilon$ is the energy, $f_d(\varepsilon)$ is the Fermi-Dirac distribution, $q_e$ is the elementary charge, and $\hbar$ is the reduced Plank constant. The first term in (A1) relates to intraband contributions of graphene, which usually dominate in the low THz range, and the second term relates to interband contributions of graphene, which become more important at higher frequencies.

From the engineering point of view, it is useful to study the surface impedance (sometimes also called complex resistivity) of a graphene monolayer defined as $Z=1/\sigma$. Fig. A1 shows the real and imaginary parts of this quantity versus different values of the chemical potential, considering the room temperature $T$ = 300 K and the relaxation time $\tau$ = 1 ps. Note that $\mu_c$ can be easily varied using an external electrostatic field.

Analysis shows that graphene behaves almost as a frequency-independent resistor, with a significant purely inductive reactance. It is observed that an increase of the chemical potential $\mu_c$ leads to the lower losses and an up-shift of the frequencies where graphene presents large inductive behavior. The latter feature makes this material appropriate for the propagation of delocalized surface plasmons, which are transverse magnetic waves

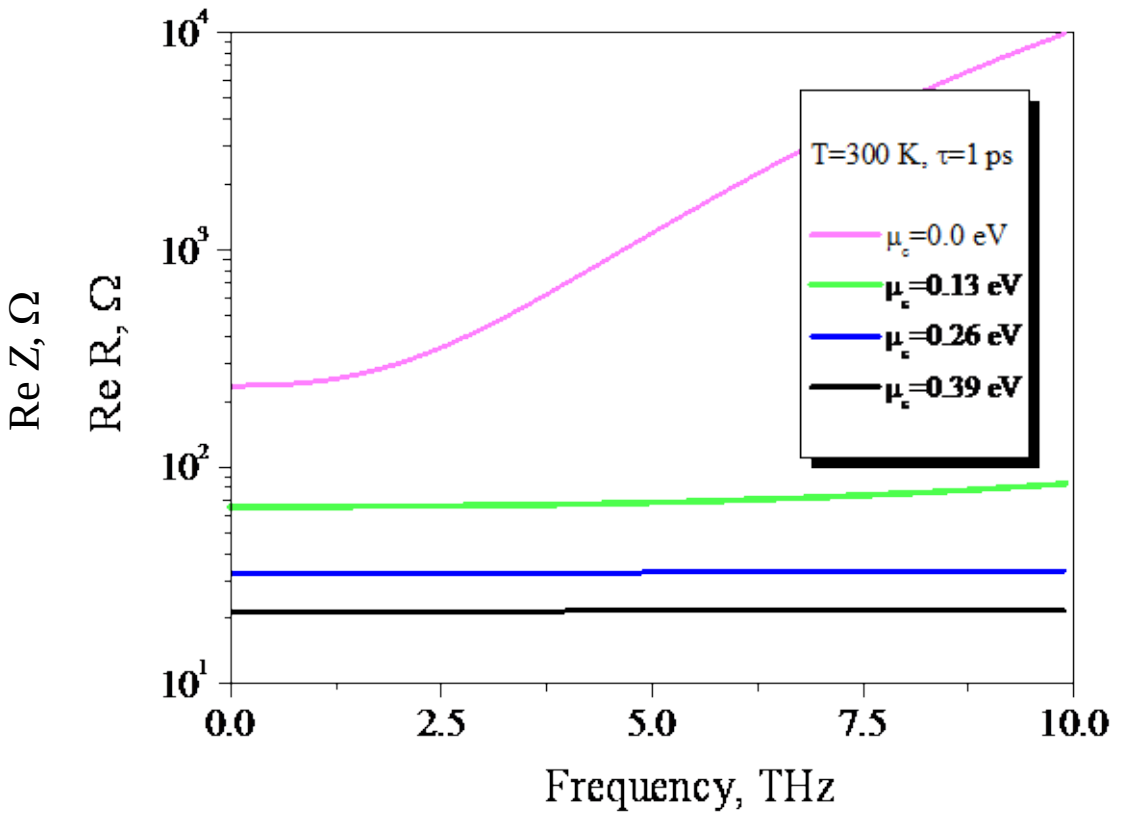

(a)

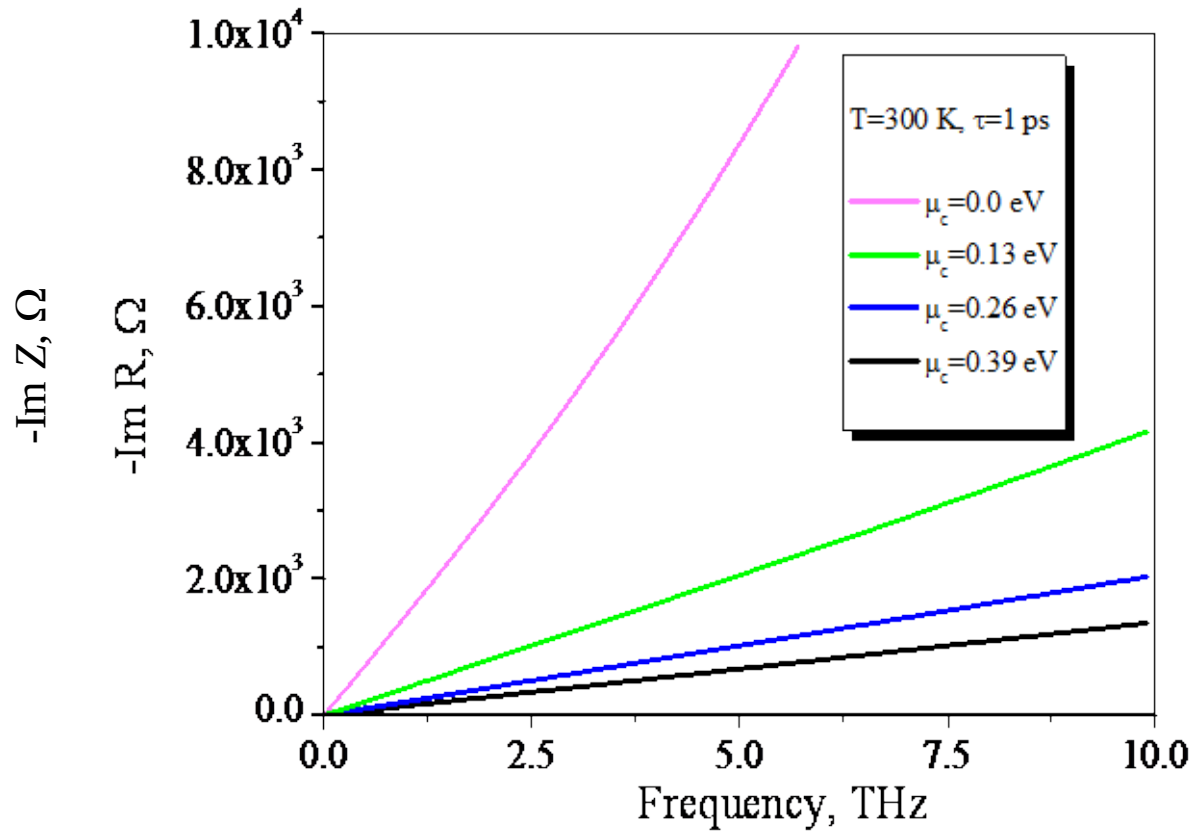

(b)

Fig. A1. Real (a) and imaginary (b) parts of the graphene impedance $Z = 1/\sigma$ in the THz range calculated at the room temperature T = 300 K and $\tau$=1 ps versus the chemical potential $\mu_c$.

traveling along the interface between graphene and dielectric [1-3].